\documentclass[12pt]{article} 
\usepackage{graphicx,floatflt,amssymb,rotate}
\usepackage{mathrsfs}
\usepackage{amssymb}
\usepackage{amsmath}
\usepackage{color}
\usepackage{graphicx}
\usepackage{subfigure}
\usepackage{multirow}
\usepackage{pstricks}
\usepackage{float}
\usepackage[toc,page]{appendix}

\begin{document}
\vskip 30pt  
 
\begin{center}  
{\Large \bf Effects of non-minimal Universal Extra Dimension on $B\rightarrow X_s\gamma$} \\
\vspace*{1cm}  
\renewcommand{\thefootnote}{\fnsymbol{footnote}}  
{ {\sf Anindya Datta$^1$ \footnote{email: adphys@caluniv.ac.in}},  
{\sf Avirup Shaw$^2$ \footnote{email: avirup.cu@gmail.com}} 
}\\  
\vspace{10pt}  
{ {\em $^1$Department of Physics, University of Calcutta,  
92 Acharya Prafulla Chandra Road, \\ Kolkata 700009, India}
\\
{\em $^2$Department of Theoretical Physics, Indian Association for the Cultivation of Science,\\
2A $\&$ B Raja S.C. Mullick Road, Jadavpur, Kolkata 700 032, India}}
\normalsize  
\end{center} 

\begin{abstract}
We estimate contributions from  Kaluza-Klein  excitations of third generation quarks and gauge bosons to the branching ratio of $B\rightarrow X_s\gamma$ decay process in 5-Dimensional Universal Extra Dimensional scenario with non-vanishing boundary localised terms. This model is conventionally known as non-minimal Universal Extra Dimensional model. We have derived the lower limit on the size of the extra dimension by comparing our theoretical estimation of the branching ratio which includes next-to-next-to leading order QCD corrections with its experimentally measured value. Coefficients of the boundary localised terms have also been constrained. 95 \% C.L. lower limit on inverse of radius of compactification ($R^{-1}$) can be as large as 670 GeV for some choice of the value of coefficients of boundary localised terms. 
\end{abstract}

\noindent PACS No: {\tt 11.10.Kk, 12.60.-i, 13.20.He}\\
\texttt{Key Words:~~Universal Extra Dimension, Kaluza-Klein, radiative decays of mesons} 
\renewcommand{\thesection}{\Roman{section}}  
\setcounter{footnote}{0}  
\renewcommand{\thefootnote}{\arabic{footnote}}

\section{Introduction}
Discovery of Higgs boson at the Large Hadron Collider (LHC) \cite{atlas, cms} has been a milestone in the history of Standard Model (SM). However, SM paradigm has not been successful in explaining various pressing issues, among them neutrino mass and mixing as well as Dark Matter (DM) problem deserve a special attention. Extension of SM with extra space-like dimensions try to cure both of these lacunae of the SM. In this article we are particularly interested in {\it SM in 4+1 dimension} which is known as Universal Extra Dimensional (UED) model \cite{acd}. 

The inclusive radiative $B$ decay processes have always been very instrumental in testing any beyond SM (BSM) scenario which (like the SM) either couples preferentially to third generation of quarks or provides extra flavour changing neutral currents (FCNCs). The world average experimental value of the branching ratio of this process is \cite{br_bsg1} 
\begin{equation}\label{br_exp}
{Br}^{exp}(B\rightarrow X_s\gamma)=(3.43\pm 0.21\pm 0.07)\times10^{-4},
\end{equation}
for photon energy $E_{\gamma} >1.6$ GeV in the $B$-meson rest frame. The corresponding SM prediction (including possible higher order corrections till-date) under the same conditions is \cite{NNLO}
\begin{equation}\label{br_sm}
{Br}^{SM}(B\rightarrow X_s\gamma)=(3.36\pm 0.23)\times10^{-4}.
\end{equation}
Thus the theoretical prediction is in good agreement with the experimental value. The small difference between central values of experimental and theoretical 
value tightly constraints  any new physics which contributes to this decay amplitude. Keeping this in mind we have calculated the branching ratio of this decay process in non-minimal UED (nmUED) model which we briefly describe below. 

In this model all the SM fields can access an extra flat space-like dimension $y$ compactified on a circle $S^1$ of radius $R$. The fields defined on this manifold are often conveniently expressed in terms of towers of 4-Dimensional (4D) Kaluza-Klein (KK) states. The zero-mode of the KK-towers is identified as the corresponding 4D SM field. A ${Z}_2$
symmetry ($y \leftrightarrow -y$) needs to be imposed to generate chiral SM fermions in the theory. Now the extra
dimension is called $S^1/Z_2$ orbifold and consequently physical domain extends
from $y = 0$ to $y = \pi R$. The $y \leftrightarrow -y$ symmetry leads to a
conserved KK-parity  $=(-1)^n$.  $n$ is defined as KK-number which represents discretised momentum along the $y$-direction. The conservation
of KK-parity ensures that the lightest Kaluza-Klein particle (LKP) with KK-number one ($n=1$) cannot decay to a pair of SM particles and is absolutely stable. Hence the LKP can be considered as a potential DM candidate of this scenario \cite{ued_dm, relic}. Furthermore, variants of this model can address other unsolved issues of SM, like gauge coupling unifications \cite{ued_uni}, neutrino mass \cite{ued_nutrino} and fermion mass hierarchy \cite{hamed} etc.

The KK-states of all particles at the $n^{th}$ KK-level have the mass, $\sqrt{(m^2+(nR^{-1})^2)}$. $m$ is the zero-mode mass (SM particle mass) which is small compared to $R^{-1}$. This implies that  UED scenario leads to a near degenerate mass spectrum at each KK-level.  Consequently, this model has very challenging phenomenology, particularly, at the colliders. Fortunately this mass degeneracy can be lifted by the virtue of radiative corrections \cite{rad_cor_georgi, rad_cor_cheng}. There are two distinct classes  of radiative corrections. The first one called  bulk corrections (which are finite and only non-zero for KK-excitations of gauge bosons) and second one is boundary localised corrections having logarithmic dependence on the cut-off scale $\Lambda$\footnote{Since UED is an effective theory characterised by a cut-off scale $\Lambda$.}.  At the two fixed boundary points ($y=0$ and $y=\pi R$) one can allow
4D kinetic, mass and other possible interaction terms for the KK-states. In fact it is natural to expect  such terms in an extra dimensional theory like UED  as counterterms for cut-off dependent loop-induced contributions. A very unique assumption has been made in the minimal
UED (mUED) models that these boundary terms are tuned in such a way that the 5-Dimensional (5D) radiative corrections exactly vanish at the cut-off scale $\Lambda$. In general this unique choice can be avoided. Without doing the actual radiative corrections one might consider kinetic, mass as well as other interaction terms localised at the fixed points to parametrise these unknown corrections. Hence this particular scenario can be termed nmUED \cite{Dvali}-\cite{ddrs1}. Strength of different boundary localised terms (BLTs) along with radius of compactification ($R$) can be treated as free parameters of this model and using various experimental inputs one can constrain these parameters. Several such phenomenological exercise have been performed in the framework of nmUED from different perspective. For example bounds on the values of the coefficients of the boundary localised terms  are obtained from the consideration of electroweak observables \cite{flacke}, $S$, $T$ and $U$ parameters \cite{delAguila_STU, flacke_STU}, relic density \cite{tommy, ddrs2}, production and decay of SM Higgs boson \cite{tirtha}, study of LHC experiments \cite{asesh_lhc1, lhc}, $R_b$ \cite{zbb}, branching ratio of  $B_s \rightarrow \mu^+ \mu^-$ \cite{bmm}, flavour changing rare top decay \cite{Dey:2016cve} and unitarity of scattering amplitudes involving KK-excitations \cite{Jha:2016sre}.

This article has been dedicated to explore branching ratio of the decay $B\rightarrow X_s\gamma$ in the nmUED model. To the best of our knowledge, branching ratio of the decay $B\rightarrow X_s\gamma$ in the framework of nmUED has not yet been presented in the literature. Aim of our investigation will be twofold. First of all, we will try to put constraints on the BLT parameters by comparing our results with the experimental value of the concerned FCNC process. And secondly, we would like to see how far one can push the lower limit on $R^{-1}$ to higher values with non-zero  BLT parameters? Another interesting part of this exercise is to check whether this lower limit of $R^{-1}$ comparable with the results derived from the above mentioned studies or not? The same exercise in the context of UED first time has been performed in several years ago \cite{buras3}. We will see that value of the lower limit on $R^{-1}$ as derived in ref.\;\cite{buras3} would change while we compare the current experimental result \cite{br_bsg1} with our theoretical estimation  which includes next-to-next-to leading order (NNLO) correction \cite{NNLO}. In view of this we will review the lower bound on {$R^{-1}$} in UED model, where strengths of BLT parameters are considered to be zero.

In the following section, we will describe the nmUED model in brief. Then we will show the calculational details in section 3. In section 4 we will present our numerical results. Finally, we conclude in section 5.

\section{A very short overview of KK-parity conserving nmUED scenario}

In this section we briefly discuss the technicalities of the nmUED model relevant for our analysis. Further details can be found in \cite{Dvali}-\cite{ddrs1}, \cite{asesh_lhc1}, \cite{zbb}, \cite{bmm}. 

The action of 5D fermionic fields with boundary localised kinetic term (BLKT) of strength $r_f$ is given by \cite{schwinn, ddrs2, bmm}:
\begin{eqnarray} 
S_{fermion} = \int d^5x \left[ \bar{\Psi}_L i \Gamma^M D_M \Psi_L 
+ r_f\{\delta(y)+\delta(y - \pi R)\} \bar{\Psi}_L i \gamma^\mu D_\mu P_L\Psi_L  
\right. \nonumber \\
\left. + \bar{\Psi}_R i \Gamma^M D_M \Psi_R
+ r_f\{\delta(y)+\delta(y - \pi R)\}\bar{\Psi}_R i \gamma^\mu D_\mu P_R\Psi_R
\right]. 
\label{factn}
\end{eqnarray}

Here $\Psi_L(x,y)$ and $\Psi_R(x,y)$ are the 5D four component Dirac spinors, which can be written in terms of two component spinors \cite{schwinn, ddrs2, bmm}:

\begin{equation} 
\Psi_L(x,y) = \begin{pmatrix}\phi_L(x,y) \\ \chi_L(x,y)\end{pmatrix}
=   \sum_n \begin{pmatrix}\phi^{(n)}_L(x) f_L^n(y) \\ \chi^{(n)}_L(x) g_L^n(y)\end{pmatrix}, 
\label{fermionexpnsn1}
\end{equation}
\begin{equation} 
\Psi_R(x,y) = \begin{pmatrix}\phi_R(x,y) \\ \chi_R(x,y) \end{pmatrix} 
=   \sum_n \begin{pmatrix}\phi^{(n)}_R(x) f_R^n(y) \\ \chi^{(n)}_R(x) g_R^n(y) \end{pmatrix}. 
\label{fermionexpnsn2} 
\end{equation}\\

The KK-wave-functions ($f_{L(R)}$ and $g_{L(R)}$) can be expressed as the following \cite{carena, flacke, ddrs2, bmm}: 
\begin{eqnarray}
f_L^n = g_R^n = N^f_n \left\{ \begin{array}{rl}
                \displaystyle \frac{\cos\left[m_{f^{(n)}} \left (y - \frac{\pi R}{2}\right)\right]}{\cos[ \frac{m_{f^{(n)}} \pi R}{2}]}  &\mbox{for $n$ even,}\\
                \displaystyle \frac{{-}\sin\left[m_{f^{(n)}} \left (y - \frac{\pi R}{2}\right)\right]}{\sin[ \frac{m_{f^{(n)}} \pi R}{2}]} &\mbox{for $n$ odd,}
                \end{array} \right.
                \label{flgr}
\end{eqnarray}
and
\begin{eqnarray}
g_L^n =-f_R^n = N^f_n \left\{ \begin{array}{rl}
                \displaystyle \frac{\sin\left[m_{f^{(n)}} \left (y - \frac{\pi R}{2}\right)\right]}{\cos[ \frac{m_{f^{(n)}} \pi R}{2}]}  &\mbox{for $n$ even,}\\
                \displaystyle \frac{\cos\left[m_{f^{(n)}} \left (y - \frac{\pi R}{2}\right)\right]}{\sin[ \frac{m_{f^{(n)}} \pi R}{2}]} &\mbox{for $n$ odd.}
                \end{array} \right.
\end{eqnarray}
$N^f_n$, normalisation constant for $n^{th}$ KK-mode, could be readily derived form orthonormality conditions \cite{ddrs2, bmm}:
\begin{equation}\label{orthonorm}
\begin{aligned}
&\left.\begin{array}{r}
                  \int_0 ^{\pi R}
dy \; \left[1 + r_{f}\{ \delta(y) + \delta(y - \pi R)\}\right]f_L^mf_L^n\\
                  \int_0 ^{\pi R}
dy \; \left[1 + r_{f}\{ \delta(y) + \delta(y - \pi R)\}\right]g_R^mg_R^n
\end{array}\right\}=&&\delta^{n m}~;
&&\left.\begin{array}{l}
                 \int_0 ^{\pi R}
dy \; f_R^mf_R^n\\
                 \int_0 ^{\pi R}
dy \; g_L^mg_L^n
\end{array}\right\}=&&\delta^{n m}~,
\end{aligned}
\end{equation}

and it is given by:
{\small
\vspace*{-.3cm}
\begin{equation}\label{norm}
N^f_n=\sqrt{\frac{2}{\pi R}}\Bigg[ \frac{1}{\sqrt{1 + \frac{r^2_f m^2_{f^{(n)}}}{4} + \frac{r_f}{\pi R}}}\Bigg].
\end{equation}
}
Mass of $n^{th}$ KK-excitation ($m_{f^{(n)}}$) satisfies the following transcendental equations \cite{carena, ddrs2, bmm}: 
\begin{eqnarray}
  \frac{r_{f} m_{f^{(n)}}}{2}= \left\{ \begin{array}{rl}
         -\tan \left(\frac{m_{f^{(n)}}\pi R}{2}\right) &\mbox{for $n$ even,}\\
          \cot \left(\frac{m_{f^{(n)}}\pi R}{2}\right) &\mbox{for $n$ odd.}
          \end{array} \right.   
          \label{fermion_mass}      
 \end{eqnarray}


Large top quark mass plays a vital role in amplifying the quantum effects in our study. So it is important  to discuss Yukawa interactions in this scenario.   The Yukawa action with boundary localised terms of strength $r_y$ is given by \cite{bmm}:  

\begin{eqnarray}
\label{yukawa}
S_{Yukawa} &=& -\int d^5 x  \Big[\lambda^5_t\;\bar{\Psi}_L\widetilde{\Phi}\Psi_R 
  +r_y \;\{ \delta(y) + \delta(y-\pi R) \}\lambda^5_t\bar{\phi_L}\widetilde{\Phi}\chi_R+\textrm{h.c.}\Big].
\end{eqnarray}

$\lambda^5_t$ is the 5D coupling of Yukawa interaction for the third generations. 
Plugging the KK-expansions for fermions (given in Eqs.\;\ref{fermionexpnsn1} and \ref{fermionexpnsn2})
in the actions given in Eq.\;\ref{factn} and Eq.\;\ref{yukawa} we can have the bi-linear terms involving  the doublet and singlet states of the quarks. The mass matrix for $n^{th}$ KK-level is as the following \cite{bmm}: 

\begin{equation}
\label{fermion_mix}
-\begin{pmatrix}
\bar{\phi_L}^{(n)} & \bar{\phi_R}^{(n)}
\end{pmatrix}
\begin{pmatrix}
m_{f^{(n)}}\delta^{nm} & m_{t} {\mathscr{I}}^{nm}_1 \\ m_{t} {\mathscr{I}}^{mn}_2& -m_{f^{(n)}}\delta^{mn}
\end{pmatrix}
\begin{pmatrix}
\chi^{(m)}_L \\ \chi^{(m)}_R
\end{pmatrix}+{\rm h.c.}.
\end{equation}
Here, $m_t$ represents the mass of SM top quark and $m_{f^{(n)}}$ are the solutions of transcendental equations given in Eq.\;\ref{fermion_mass}. The overlap integrals (${\mathscr{I}}^{nm}_1$ and ${\mathscr{I}}^{nm}_2$) are given by \cite{bmm}:
  \[ {\mathscr{I}}^{nm}_1=\left(\frac{1+\frac{r_f}{\pi R}}{1+\frac{r_y}{\pi R}}\right)\times\int_0 ^{\pi R}\;dy\;
\left[ 1+ r_y \{\delta(y) + \delta(y - \pi R)\} \right] g_{R}^m f_{L}^n,\] \;\;{\rm and}\;\;\[{\mathscr{I}}^{nm}_2=\left(\frac{1+\frac{r_f}{\pi R}}{1+\frac{r_y}{\pi R}}\right)\times\int_0 ^{\pi R}\;dy\;
 g_{L}^m f_{R}^n .\]

For both the cases of $n=m$ and $n\neq m$ the integral ${\mathscr{I}}^{nm}_1$ is non zero. But for $r_y = r_f$, this integral equal to 1 (when $n =m$) or 0 ($n \neq m$). And the integral ${\mathscr{I}}^{nm}_2$ is non zero only when  $n =m$ and equal to 1 in the limit $r_y = r_f$. One should note that, in our analysis we choose an equality condition ($r_y$=$r_f$) just to avoid the complicacy of mode mixing and construct a simpler form of fermion mixing matrix \cite{zbb, bmm}. With this motivation, in the rest of our analysis we will stick to the choice of equal $r_y$ and $r_f$\footnote{However, in general one can take unequal strengths of boundary terms for Yukawa and kinetic interaction for fermions.}. 

After imposing the above equality condition the resulting mass matrix (given in Eq.\;\ref{fermion_mix}) can easily be diagonalised by following bi-unitary transformations for the left- and right-handed fields respectively \cite{bmm}:
\begin{equation}
U_{L}^{(n)}=\begin{pmatrix}
\cos\alpha_{tn} & \sin\alpha_{tn} \\ -\sin\alpha_{tn} & \cos\alpha_{tn}
\end{pmatrix},~~U_{R}^{(n)}=\begin{pmatrix}
\cos\alpha_{tn} & \sin\alpha_{tn} \\ \sin\alpha_{tn} & -\cos\alpha_{tn}
\end{pmatrix},
\end{equation}
where $\alpha_{tn}[ = \frac12\tan^{-1}\left(\frac{m_{t}}{m_{f^{(n)}}}\right)]$ is the mixing angle. The gauge eigen states $\Psi_L(x,y)$ and $\Psi_R(x,y)$ and mass eigen states $T^1_t$ and $T^2_t$ are related by the following relations \cite{bmm}:
%
\begin{tabular}{p{8cm}p{8cm}}
{\begin{align}
&{\phi^{(n)}_L} =  \cos\alpha_{tn}T^{1(n)}_{tL}-\sin\alpha_{tn}T^{2(n)}_{tL},\nonumber \\
&{\chi^{(n)}_L} =  \cos\alpha_{tn}T^{1(n)}_{tR}+\sin\alpha_{tn}T^{2(n)}_{tR},\nonumber
\end{align}}
&
{\begin{align}
&{\phi^{(n)}_R} =  \sin\alpha_{tn}T^{1(n)}_{tL}+\cos\alpha_{tn}T^{2(n)}_{tL},\nonumber \\
&{\chi^{(n)}_R} =  \sin\alpha_{tn}T^{1(n)}_{tR}-\cos\alpha_{tn}T^{2(n)}_{tR}.
\end{align}}
\end{tabular}
%
The mass eigen value at $n^{th}$ KK-level is $M_{t^{(n)}} \equiv \sqrt{m_{t}^{2}+m^2_{f^{(n)}}}$. This is same for both physical eigen states $T^{1(n)}_t$ and $T^{2(n)}_t$. 

Let us focus on the necessary interactions of gauge field $W^i_M(\equiv W^i_\mu, W^i_4$)\footnote{$i$, is the $SU(2)_L$ group index, runs from 1 to 3.} and scalar field $\Phi$. The 5D kinetic actions with their respective boundary localised terms are given by \cite{flacke, zbb, bmm}:
\begin{eqnarray}
\label{gauge}
S^W_{gauge} &=& -\frac{1}{4}\int d^5 x \Big[W^{MNi}W_{MN}^{i}+r_V\{ \delta(y) + \delta(y - \pi R)\}W^{\mu \nu i}W_{\mu \nu}^{i} \Big],\\
\label{higgs}
S_{scalar} &=& \int d^5 x  \Big[\left(D^{M}\Phi\right)^{\dagger}\left(D_{M}\Phi\right) + r_{\phi}\{ \delta(y) + \delta(y - \pi R)\}\left(D^{\mu}\Phi\right)^{\dagger}\left(D_{\mu}\Phi\right) \Big],
\end{eqnarray}
where, $r_V$ and $r_\phi$ denote the strengths of the BLKTs for gauge and scalar fields respectively. KK-expansion of the above mentioned fields are given as follows  \cite{bmm}:
\begin{equation}\label{W}
W^i_{\mu}(x,y)=\sum_n W_{\mu}^{i(n)}(x) a^n(y)~;~~
W^i_{4}(x,y)=\sum_n W_{4}^{i(n)}(x) b^n(y),
\end{equation}
and
\begin{equation}\label{phi}
\Phi(x,y)=\sum_n \Phi^{(n)}(x) h^n(y).
\end{equation}
An important issue in the context of the present work is gauge fixing action/mechanism, as we will compute relevant loop diagrams in Feynman gauge.  The gauge fixing\footnote{A comprehensive study on gauge fixing action/mechanism in nmUED can be found in ref.\;\cite{gf}.} action in nmUED scenario can be written as \cite{zbb, bmm, gf}:
\begin{eqnarray}
\label{gauge_fix}
S_{\rm gf}^{W} &=& -\frac{1}{\xi _y}\int d^5x\Big\vert\partial_{\mu}W^{\mu +}+\xi_{y}(\partial_{y}W^{4+}+iM_{W}\phi^{+}\{1 + r_{V}\left( \delta(y) + \delta(y - \pi R)\right)\})\Big \vert ^2 .\nonumber \\
\end{eqnarray}
$\xi _y$ is related with {\em physical} gauge fixing parameter $\xi$ (with values 0 (Landau gauge), 1 (Feynman gauge) or $\infty$ (Unitary gauge)) through \cite{zbb, bmm, gf}, 
\begin{equation}
\xi= \xi_y \{1 + r_{V}\left( \delta(y) + \delta(y - \pi R)\right)\}.
\end{equation}
 
It is necessary to set $r_V=r_\phi$ for proper gauge fixing \cite{zbb, bmm, gf}. As a result KK-masses for gauge and scalar fields are equal ($m_{V^{(n)}}(=m_{\phi^{(n)}})$) and satisfy the same transcendental equation (Eq.\;\ref{fermion_mass}). Mass eigen value of $n^{th}$ KK-mode of gauge fields ($W^{\mu (n)\pm}$) and charged Higgs ($H^{(n)\pm}$) is $M_{W^{(n)}}=\sqrt{M_{W}^{2}+m^2_{V^{(n)}}}$.  The mass of Goldstone bosons ($G^{(n)\pm}$) corresponding to the gauge fields $W^{\mu (n)\pm}$ has the same value $M_{W^{(n)}}$ in 't-Hooft Feynman gauge \cite{zbb, bmm}.  

Necessary interactions for our calculation can be derived by integrating out the 5-D action over the extra space-like dimension  after substituting the $y$-dependent function for the respective fields in 5-D action\footnote{As a result some of the interactions are modified by so called overlap integrals. The expressions of the overlap integrals have been given in appendix \ref{fyerul}.}. We list necessary Feynman rules in appendix \ref{fyerul}. For further details we refer the reader to \cite{bmm}.

\section{\boldmath{$B\rightarrow X_s\gamma$} in nmUED}
Radiative decay of $B$ meson is sensitive to any new physics that preferentially couples to third generation of quarks. Several experimental collaboration (CLEO~\cite{Chen:2001fj}, Belle~\cite{Abe:2001hk,Limosani:2009qg}, and
BABAR~\cite{Lees:2012ym}-\cite{Aubert:2007my}) have been looking for the signals of the decay process $B\to X_s\gamma$ for the last couple of decades. In this section
we will presents details of the calculation of the branching ratio of $B\to X_s\gamma$ in nmUED model. nmUED model like SM has only one Higgs doublet, so there is no FCNC that can generate a chirality flip. Therefore,  in this model leading order (LO) contributions to dipole operators are one loop suppressed as in SM. However, there are more one loop diagrams than in SM, due to large number of KK-particles which have to be taken into account. Hence, to evaluate the total contributions to the dipole operators, we just simply add the KK-contributions to that of SM. Following the same strategy of the ref.\;\cite{buras3} at first we will find Wilson coefficients of dipole operators at the LO level. Finally, utilising the technique of ref.\;\cite{NNLO} we will estimate branching ratio of $B\to X_s\gamma$ incorporating NNLO corrections. 

\subsection{Effective Hamiltonian for $B\rightarrow X_s\gamma$ decay at leading order}
 
$B\to X_s\gamma$ decay is indicated at quark level by $b\to s\gamma$ transition. The effective Hamiltonian for this transition at scale $\mu_b={\cal O}(m_b)$\footnote{$m_b$ is the mass of bottom quark.}
can be written in the following form (see ref.\;\cite{bsg_rvw, Buras:1997fb}):
\begin{equation} \label{Heff_at_mu}
{\cal H}_{\rm eff}(b\to s\gamma) = - \frac{G_{\rm F}}{\sqrt{2}} V_{ts}^* V_{tb}
\left[ \sum_{i=1}^6 C_i(\mu_b) Q_i + C_{7\gamma}(\mu_b) Q_{7\gamma}
+C_{8G}(\mu_b) Q_{8G} \right]\,.
\end{equation}
$G_F$ denotes the Fermi constant and $V_{ij}$ are the Cabibbo-Kobayashi-Maskawa (CKM) matrix elements. $Q_1....Q_6$ are the local operators which represent four quark interactions. Explicit form of these operators can be found in \cite{bsg1_6}. The residual $Q_{7\gamma}$ (electromagnetic dipole) and $Q_{8G}$ (chromomagnetic dipole) are the most important operators for this decay as follows:
\begin{equation}\label{O6B}
Q_{7\gamma}  =  \frac{e}{8\pi^2} m_b \bar{s}_\alpha \sigma^{\mu\nu}
          (1+\gamma_5) b^\alpha F_{\mu\nu},\qquad            
Q_{8G}     =  \frac{g_s}{8\pi^2} m_b \bar{s}^\alpha \sigma^{\mu\nu}
   (1+\gamma_5)T^a_{\alpha\beta} b^\beta G^a_{\mu\nu}\;, 
\end{equation}
with $\sigma^{\mu\nu}=\frac{i}{2}[\gamma^\mu, \gamma^\nu]$. $T^a$ are the generators of $SU(3)_C$ gauge group. The Wilson coefficients $C_i(\mu_b)$ have been evolved from the electroweak scale down to $\mu_b = m_b$ through renormalisation group (RG) equations \cite{buras3, bsg_rvw, Buras:1997fb}.
These coefficients ($C_{7\gamma}(\mu_b)$ and $C_{8G}(\mu_b)$) at the LO level are given by the following relations:

\begin{eqnarray}
\label{C7eff}
C_{7\gamma}^{(0)eff}(\mu_b) & = & 
\eta^\frac{16}{23} C_{7\gamma}^{(0)}(\mu_W) + \frac{8}{3}
\left(\eta^\frac{14}{23} - \eta^\frac{16}{23}\right) C_{8G}^{(0)}(\mu_W) +
 C_2^{(0)}(\mu_W)\sum_{i=1}^8 h_i \eta^{a_i},
\\
\label{C8eff}
C_{8G}^{(0)eff}(\mu_b) & = & 
\eta^\frac{14}{23} C_{8G}^{(0)}(\mu_W) 
   + C_2^{(0)}(\mu_W) \sum_{i=1}^8 \bar h_i \eta^{a_i},
\end{eqnarray}
with
\begin{equation}
\eta  =  \frac{\alpha_s(\mu_W)}{\alpha_s(\mu_b)},~~~\alpha_s(\mu_b) = \frac{\alpha_s(M_Z)}{1 
- \beta_0 \frac{\alpha_s(M_z)}{2\pi} \, \ln(M_Z/\mu_b)}, \qquad 
\beta_0=\frac{23}{3}~,
\label{eq:asmumz}
\end{equation}
and 
\begin{eqnarray}\label{c2}
C^{(0)}_2(\mu_W) &=& 1,\\                             
C^{(0)}_{7\gamma} (\mu_W) &=& -\frac{1}{2} D'(x_t, r_f, r_V, R^{-1}),\\ 
C^{(0)}_{8G}(\mu_W) &=& -\frac{1}{2} E'(x_t, r_f, r_V, R^{-1}).
\end{eqnarray}
Rest of the Wilson coefficients vanish at electroweak scale ($\mu_W$). 
The LO approximation is indicated by the superscript \textquotedblleft0\textquotedblright. The values of $a_i$, $h_i$ and $\bar h_i$ can be obtained from \cite{buras3}.

The functions $D'(x_t, r_f, r_V, R^{-1})$ and $E'(x_t, r_f, r_V, R^{-1})$ are the total (SM+nmUED) contributions at the LO as given by:

\begin{equation}
D'(x_t, r_f, r_V, R^{-1})=D'_0(x_t)+
\sum_{n=1}^\infty D'_n(x_t,x_{f^{(n)}},x_{V^{(n)}}),
\label{dprime_sum}
\end{equation}
and 
\begin{equation}
E'(x_t, r_f, r_V, R^{-1})=E'_0(x_t)+
\sum_{n=1}^\infty E'_n(x_t,x_{f^{(n)}},x_{V^{(n)}}).
\label{eprime_sum}
\end{equation}

Here $D'_0(x_t)$ and $E'_0(x_t)$ are the SM contributions at the electroweak scale \cite{lim}:

\begin{equation}
D'_0(x_t)= -{{(8x_t^3 + 5x_t^2 - 7x_t)}\over{12(1-x_t)^3}}+ 
          {{x_t^2(2-3x_t)}\over{2(1-x_t)^4}}\ln x_t~,
\end{equation}
\begin{equation}
E'_0(x_t)=-{{(x_t^3-5x_t^2-2x_t)}\over{4(1-x_t)^3}} + {3\over2}
{{x_t^2}\over{(1 - x_t)^4}} \ln x_t~,
\end{equation}
where $x_t=\frac{m^2_t}{M^2_W}$, $x_{V^{(n)}}=\frac{m^2_{V^{(n)}}}{M^2_W}$ and $x_{f^{(n)}}=\frac{m^2_{f^{(n)}}}{M^2_W}$. $m_{V^{(n)}}$ and $m_{f^{(n)}}$ are the solutions of transcendental equation given in Eq.\;\ref{fermion_mass}.

\begin{figure}[H]
\begin{center}
\includegraphics[scale=0.82,angle=0]{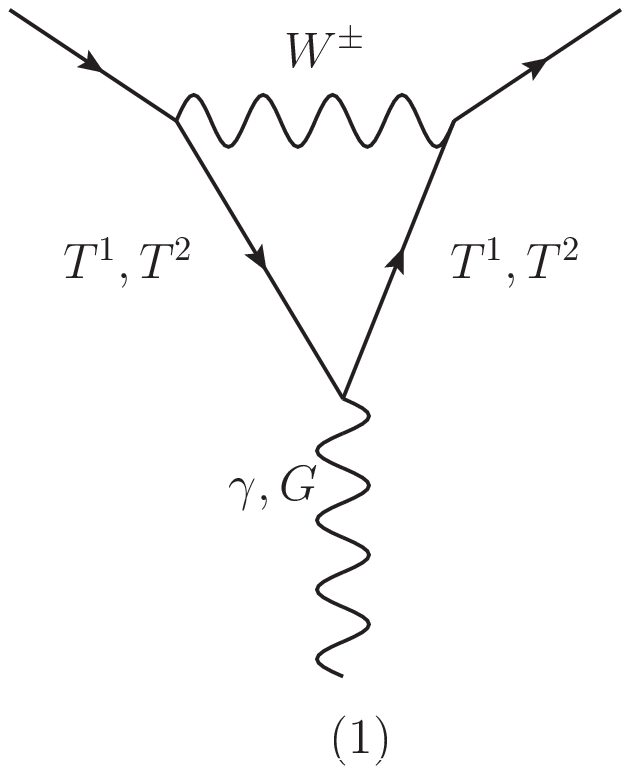}
\includegraphics[scale=0.82,angle=0]{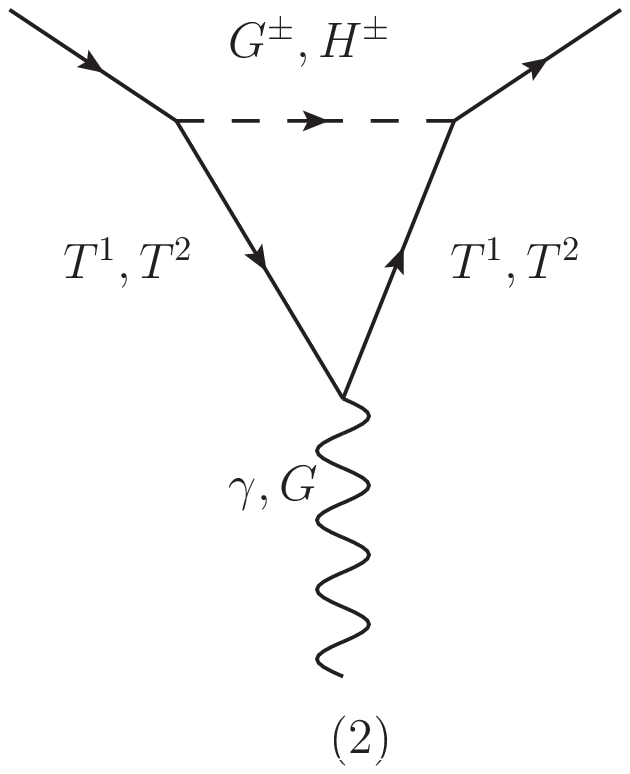}
\includegraphics[scale=0.82,angle=0]{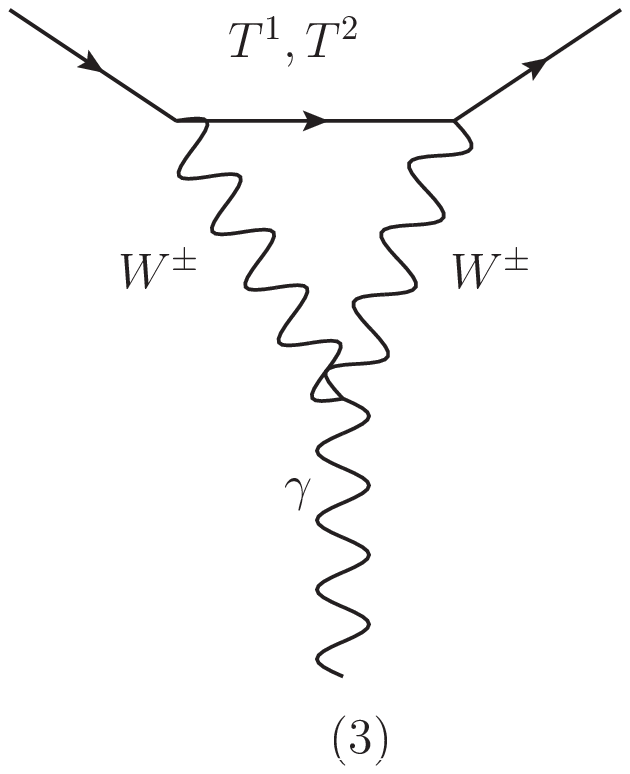}
\includegraphics[scale=0.82,angle=0]{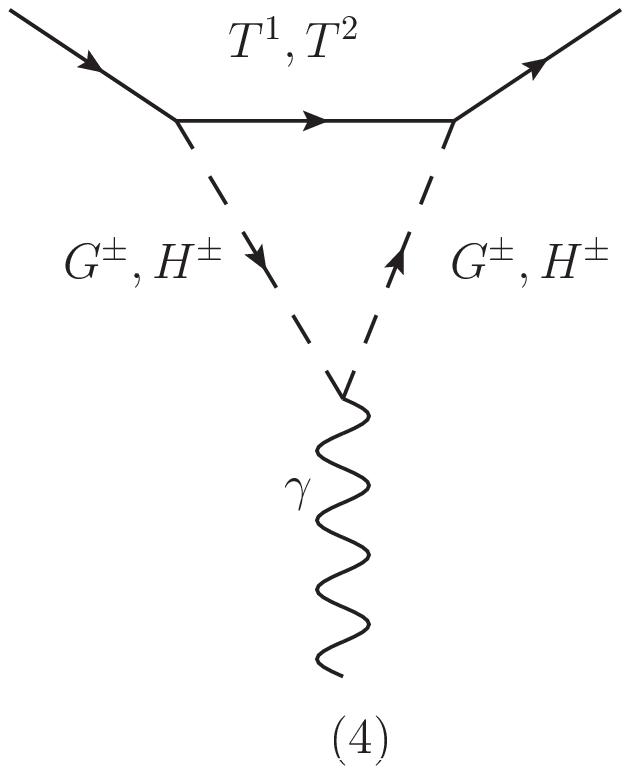}
\includegraphics[scale=0.82,angle=0]{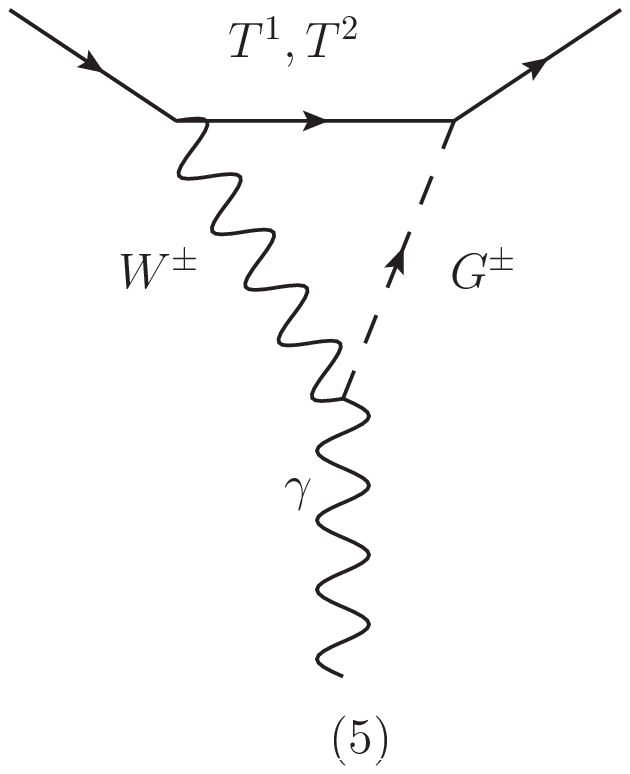}
\includegraphics[scale=0.82,angle=0]{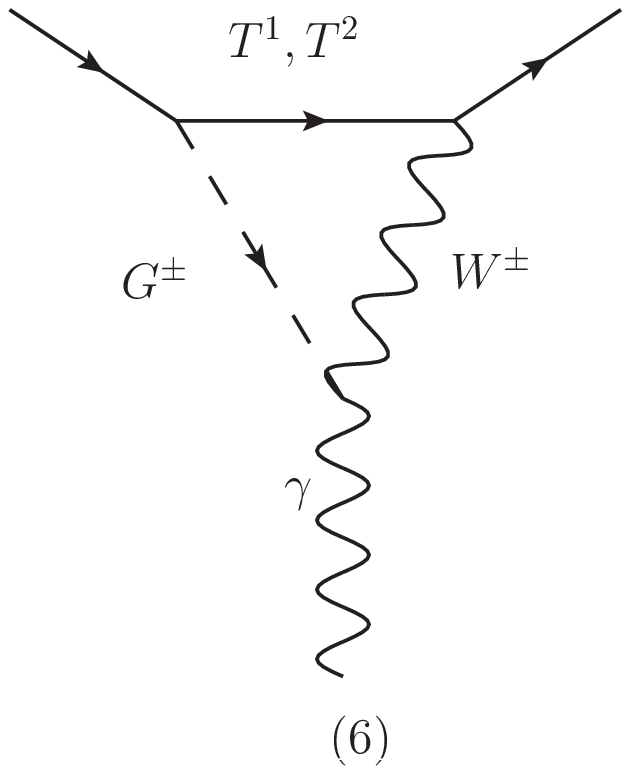}
\caption{Magnetic penguin diagrams}
\label{magnetic_pen}
\end{center}
\end{figure}

We will now present the nmUED contribution to the magnetic penguin diagrams. But before delving into that we must mention an important issue. Due to the presence 
 several BLTs in the nmUED action, KK-masses and  couplings involving KK-excitations are non-trivially modified in comparison to their UED counterparts. 
 Consequently, it would not be possible to get the expressions of $D'$ and $E'$ in nmUED simply by rescaling the results of UED model \cite{buras3}. Hence, 
 we have computed the functions $D'_n(x_t,x_{f^{(n)}},x_{V^{(n)}})$ and $E'_n(x_t,x_{f^{(n)}},x_{V^{(n)}})$ starting from the scratch. One can readily see (from Eqs.\;\ref{dprime} 
 and \ref{eprime}) that they are drastically different from that of the UED version (Eqs.\;3.33 and 3.34 of ref.\;\cite{buras3}). However, if we switch-off the boundary 
 terms i.e., setting $r_f, r_V = 0$, we can reproduce the expressions given by the Eqs.\;3.33 and 3.34 in ref.\;\cite{buras3} from our results. The functions 
 $D'_n(x_t,x_{f^{(n)}},x_{V^{(n)}})$ and $E'_n(x_t,x_{f^{(n)}},x_{V^{(n)}})$ are the sum of the KK-contributions that are calculated from the magnetic penguin diagrams (given in Fig.\;\ref{magnetic_pen}) in nmUED model  with on shell photon and gluon respectively.

While we compute the one loop penguin diagrams to estimate the contributions of KK-excitation to the branching ratio of $B\rightarrow X_s\gamma$, we have 
considered interactions which couple a zero-mode field to a pair of KK-excitations having same KK-number. Since we have explicitly checked that the final results 
would not change significantly even if one sums all the possible off-diagonal contributions\footnote{In nmUED, one can have non-zero interactions involving 
KK-excitations with  KK-numbers $n, m~{\rm and}~p$ where $n+m+p$ is an even number. This is a direct consequences of KK-parity conservation.} \cite{zbb, bmm}. 

\newpage
Let us write down the function $D'_n(x_t,x_{f^{(n)}},x_{V^{(n)}})$ which is obtained from electromagnetic magnetic penguin diagrams as in Fig.\;\ref{magnetic_pen},
{\small
\begin{eqnarray}\label{dprime}
D'_n(x_t,x_{f^{(n)}},x_{V^{(n)}})&=&\frac23E'_n(x_t,x_{f^{(n)}},x_{V^{(n)}})
+\frac{(I^n_1)^2}{12}\Bigg[\frac{1}{(1+x_{V^{(n)}}-x_{f^{(n)}})^4}\Bigg({\bigg(-31+18\ln(\frac{x_{f^{(n)}}}{1+x_{V^{(n)}}})\bigg)x^3_{f^{(n)}}}
\nonumber \\ &&
-{3\bigg(-19+2\ln(\frac{x_{f^{(n)}}}{1+x_{V^{(n)}}})\bigg)x^2_{f^{(n)}}(1+x_{V^{(n)}})}
{-33x_{f^{(n)}}(1+x_{V^{(n)}})^2+7(1+x_{V^{(n)}})^3}\Bigg)
\nonumber \\ &&
-\frac{1}{(1-x_t+x_{V^{(n)}}-x_{f^{(n)}})^4}\Bigg({6\ln(\frac{x_{f^{(n)}}+x_t}{1+x_{V^{(n)}}})(x_{f^{(n)}}+x_t)^2(-1+3x_{f^{(n)}}+3x_t-x_{V^{(n)}})}
\nonumber \\ &&
+\bigg({12(x_{f^{(n)}}+x_t)^2+12(x_{f^{(n)}}+x_t)(-1+x_t+x_{f^{(n)}}-x_{V^{(n)}})}
\nonumber \\ &&
+7(-1+x_t+x_{f^{(n)}}-x_{V^{(n)}})^2\bigg)
{(-1+x_t+x_{f^{(n)}}-x_{V^{(n)}})}\Bigg)\Bigg]
\nonumber \\ &&
+\frac{1}{12}\Bigg[\frac{(I^n_2)^2}{(1+x_{V^{(n)}}-x_{f^{(n)}})^4}\bigg(-2x^3_{f^{(n)}}+3\left(-1+2\ln(\frac{x_{f^{(n)}}}{1+x_{V^{(n)}}})\right)x^2_{f^{(n)}}(1+x_{V^{(n)}})
\nonumber \\ &&
+6x_{f^{(n)}}(1+x_{V^{(n)}})^2-(1+x_{V^{(n)}})^3\bigg)
-\frac{1}{(1-x_t+x_{V^{(n)}}-x_{f^{(n)}})^4}
\nonumber \\ &&
\Bigg((I^n_1)^2x_t\bigg(-6\ln(\frac{x_{f^{(n)}}+x_t}{1+x_{V^{(n)}}})(x_{f^{(n)}}+x_t)(1+x_{V^{(n)}})
\nonumber \\ &&
(x_{f^{(n)}}+x_t-2(1+x_{V^{(n)}}))
+(1-x_t-x_{f^{(n)}}+x_{V^{(n)}})
\nonumber \\ &&
\left(-4x^2_{f^{(n)}}-4x^2_t+5x_t(1+x_{V^{(n)}})+5(1+x_{V^{(n)}})^2+x_{f^{(n)}}(5-8x_t+5x_{V^{(n)}})\right)\bigg)
\nonumber \\ &&
+(I^n_2)^2\bigg(6\ln(\frac{x_{f^{(n)}}+x_t}{1+x_{V^{(n)}}})(x_{f^{(n)}}+x_t)^2(1+x_{V^{(n)}})+(1-x_t-x_{f^{(n)}}+x_{V^{(n)}})
\nonumber \\ &&
\left(2x^2_{f^{(n)}}+2x^2_t+5x_t(1+x_{V^{(n)}})-(1+x_{V^{(n)}})^2+x_{f^{(n)}}(5+4x_t+5x_{V^{(n)}})\right)\bigg)
\Bigg)\Bigg]
\nonumber \\ &&
+\frac{(I^n_1)^2}{4}\Bigg[\frac{1}{(1-x_t+x_{V^{(n)}}-x_{f^{(n)}})^3}\bigg(2\ln(\frac{x_{f^{(n)}}+x_t}{1+x_{V^{(n)}}})(x_{f^{(n)}}+x_t)^2
\nonumber \\ &&
-(-1+x_t+x_{f^{(n)}}-x_{V^{(n)}})(-1+3x_t+3x_{f^{(n)}}-x_{V^{(n)}})\bigg)
\nonumber \\ &&
+\frac{1}{(1+x_{V^{(n)}}-x_{f^{(n)}})^3}\bigg(\left(3-2\ln(\frac{x_{f^{(n)}}}{1+x_{V^{(n)}}})\right)x^2_{f^{(n)}}
\nonumber \\ &&
-4x_{f^{(n)}}(1+x_{V^{(n)}})+(1+x_{V^{(n)}})^2\bigg)\Bigg],
\end{eqnarray}
}

while the function $E'_n(x_t,x_{f^{(n)}},x_{V^{(n)}})$ is related to the first two diagrams (given in Fig.\;\ref{magnetic_pen}) only when we consider the chromomagnetic operator in the $b\rightarrow s G$ transition. The form of the function $E'_n(x_t,x_{f^{(n)}},x_{V^{(n)}})$ in nmUED is given as the following:

\begin{eqnarray}\label{eprime}
E'_n(x_t,x_{f^{(n)}},x_{V^{(n)}})&=& \frac{1}{12}\Bigg[ \frac{6\bigg(2(I^n_1)^2+\left(1+2\ln(\frac{x_{f^{(n)}}}{1+x_{V^{(n)}}})\right)(I^n_2)^2\bigg)x^2_{f^{(n)}}}{(-1+x_{f^{(n)}}-x_{V^{(n)}})^3}-\frac{8(I^n_1)^2-2(I^n_2)^2}{(-1+x_{f^{(n)}}-x_{V^{(n)}})} 
\nonumber \\ &&
+\frac{6\ln(\frac{x_{f^{(n)}}+x_t}{1+x_{V^{(n)}}})(x_{f^{(n)}}+x_t)^3((I^n_2)^2+(I^n_1)^2(2+x_t))}{(-1+x_t+x_{f^{(n)}}-x_{V^{(n)}})^4}
-\frac{6(x_{f^{(n)}}+x_t)^2}{(-1+x_t+x_{f^{(n)}}-x_{V^{(n)}})^3}
\nonumber \\ &&
\bigg(\left(1+2\ln(\frac{x_{f^{(n)}}+x_t}{1+x_{V^{(n)}}})\right)(I^n_2)^2+(I^n_1)^2\left(2+\left(1+4\ln(\frac{x_{f^{(n)}}+x_t}{1+x_{V^{(n)}}})\right)x_t\right)\bigg)
\nonumber \\ &&
+\frac{3(x_{f^{(n)}}+x_t)}{(-1+x_t+x_{f^{(n)}}-x_{V^{(n)}})^2}\Bigg(\left(3+2\ln(\frac{x_{f^{(n)}}+x_t}{1+x_{V^{(n)}}})\right)(I^n_2)^2
\nonumber \\ &&
+(I^n_1)^2\left(-2-4\ln(\frac{x_{f^{(n)}}+x_t}{1+x_{V^{(n)}}})+\left(7+10\ln(\frac{x_{f^{(n)}}+x_t}{1+x_{V^{(n)}}})\right)x_t\right)\Bigg)
\nonumber \\ &&
-\frac{2\bigg((I^n_2)^2+2(I^n_1)^2\left(-2+\left(5+3\ln(\frac{x_{f^{(n)}}+x_t}{1+x_{V^{(n)}}})\right)x_t\right)\bigg)}{(-1+x_t+x_{f^{(n)}}-x_{V^{(n)}})}
\nonumber \\ &&
-\frac{6\ln(\frac{x_{f^{(n)}}}{1+x_{V^{(n)}}})(2(I^n_1)^2+(I^n_2)^2)x^3_{f^{(n)}}}{(-1+x_{f^{(n)}}-x_{V^{(n)}})^4}
\nonumber \\ && 
-\frac{3x_{f^{(n)}}\bigg(\left(-2-4\ln(\frac{x_{f^{(n)}}}{1+x_{V^{(n)}}})\right)(I^n_1)^2+(I^n_2)^2\left(3+2\ln(\frac{x_{f^{(n)}}}{1+x_{V^{(n)}}})\right)\bigg)}{(-1+x_{f^{(n)}}-x_{V^{(n)}})^2}\Bigg].
\end{eqnarray}

Expressions for $I^n_1$ and $I^n_2$ are  given in the appendix \ref{fyerul} (see Eqs.\;\ref{i1} and \ref{i2}).

\subsection{Branching fraction for $B\rightarrow X_s\gamma$ decay}
Radiative decay of $B$ meson shows strong dependence on {\it b} quark mass ($m_b$) and the CKM matrix ($V_{\rm CKM}$) elements. In order to reduce the uncertainties on $m_b$ and $V_{\rm CKM}$ it is a usual practice to normalise it
by the measured semileptonic decay rate $Br(B \to X_c e \bar{\nu}_e)$. Finally, in the leading logarithmic approximation one can write this ratio as: 
\begin{equation}\label{main}
\frac{Br(B \rightarrow X_s \gamma)}
     {Br(B \rightarrow X_c e \bar{\nu}_e)}=
 \frac{|V_{ts}^* V_{tb}^{}|^2}{|V_{cb}|^2} 
\frac{6 \alpha_{\rm em}}{\pi f(z)}~|C^{(0){\rm eff}}_{7}(\mu_b)|^2\,,
\end{equation}

where,
\begin{equation}\label{phase}
f(z)=1-8z^2+8z^6-z^8-24z^4\ln z~,
\end{equation}
 
is known as the phase space factor in $Br(B \to X_c e \bar{\nu}_e)$ with $z=m_c/m_b$. $m_c$ being the charm quark mass and $\alpha_{\rm em}$ is the fine structure constant.

$B\rightarrow X_s\gamma$ branching ratio has been predicted in the SM at a very high level of accuracy with the incorporation of  higher order QED and QCD corrections.  As for example refs.\;\cite{nlo_mm, nuebrt} provide the full next-to leading order (NLO) QCD and QED corrections in two different ways. Accuracy of the present experimental data demands that we should also include NNLO QCD corrections in our analysis. The first attempt to estimate NNLO QCD corrections for this process in SM was presented in ref.\;\cite{nnlo_2006}. Finally a recent article \cite{NNLO} provides an updated and more complete NNLO QCD corrections to this process. Following \cite{NNLO} one can incorporate NNLO QCD corrections to the branching ratio of $B\rightarrow X_s\gamma$ in a model of BSM, provided the BSM contributions are additive to the SM Wilson coefficients. In the following, we simply follow the procedure of ref.\;\cite{NNLO}:
\begin{equation}\label{nnlo}
{Br}^{\rm NNLO}(B\rightarrow X_s\gamma)\times10^{4}=(3.36\pm 0.23) -8.22\Delta C_7-1.99\Delta C_8.
\end{equation}
Here $\Delta C_7$ and $\Delta C_8$ stand for  the BSM contributions to Wilson coefficients for electromagnetic and chromomagnetic 
dipole operators. According to our convention, $\Delta C_7=-\frac12\sum_{n=1}^\infty D'_n(x_t,x_{f^{(n)}},x_{V^{(n)}})$ and $\Delta C_8=-\frac12\sum_{n=1}^\infty E'_n(x_t,x_{f^{(n)}},x_{V^{(n)}})$. In the 
following section we will present the numerical estimates of  the branching fraction  in nmUED models including NNLO QCD corrections.

\section{Numerical results}
The branching ratio of  $B\rightarrow X_s\gamma$ depends on several Wilson coefficients. Among which, the coefficients of chromomagnetic dipole and electromagnetic dipole operators ($D'_n(x_t,x_{f^{(n)}},x_{V^{(n)}})$ and $E'_n(x_t,x_{f^{(n)}},x_{V^{(n)}})$) depend on fermion as well as gauge boson KK-masses, SM $W$-boson mass $M_W$ and top quark mass $m_t$\footnote{We have used $M_W=80.38$ GeV and  $m_t=173.21$ GeV as given in ref.\;\cite{pdg}.}. In view of the effect of SM Higgs mass on vacuum stability in mUED model \cite{kksum} we sum the contributions up to 5 KK-levels\footnote{Earlier articles used 20-30 KK-levels while summing up the contributions from KK-modes.} while estimating Wilson coefficients which are added to SM counterparts. This sum is convergent in UED model with one extra space-like dimension, as long as we confine ourselves to one loop calculation \cite{gg_converge}. 


\subsection{Probable bounds on $R^{-1}$ in nmUED scenario}
\begin{figure}[t]
\begin{center}
\includegraphics[scale=1,angle=0]{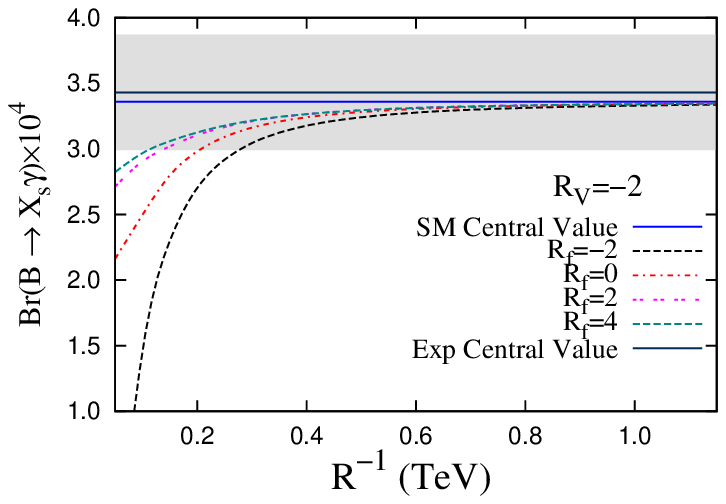}
\includegraphics[scale=1,angle=0]{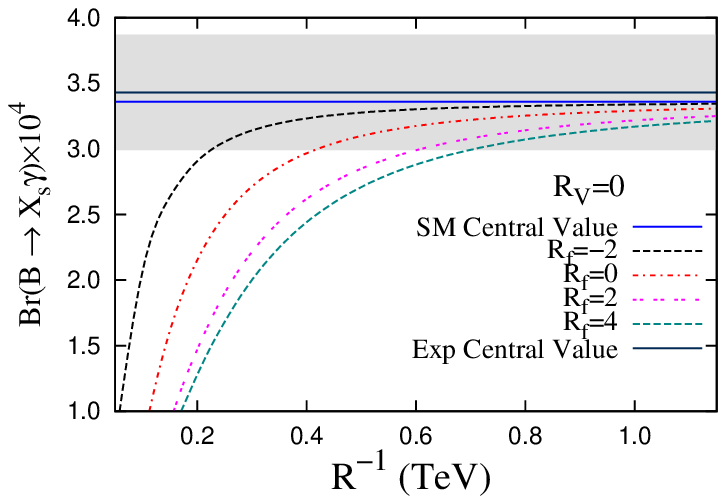}
\includegraphics[scale=1,angle=0]{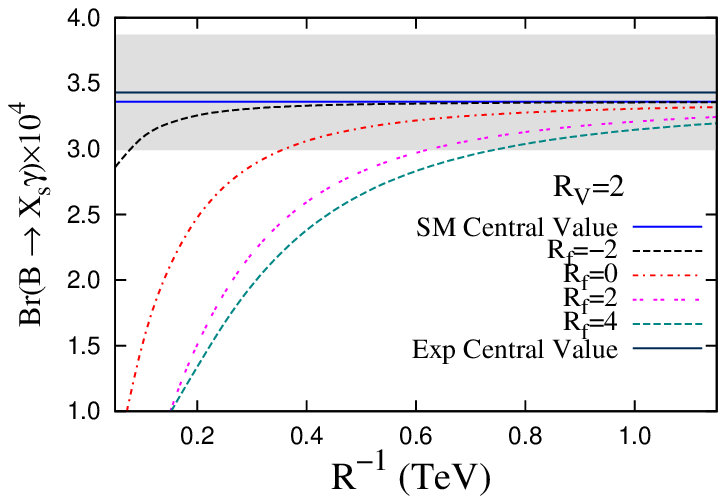}
\includegraphics[scale=1,angle=0]{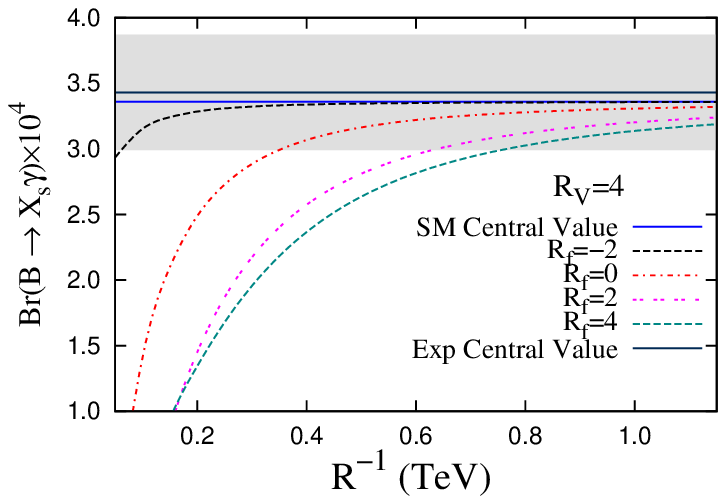}
\caption{Dependence of the branching ratio ($B\rightarrow X_s\gamma$) on $R^{-1}$ for several values of $R_f=r_f/R$. The four panels correspond to different $R_V=r_V/R$. We have used the central value in Eq.\;\ref{nnlo} to generate the curves for different values of gauge and fermion BLT parameters. We sum the contributions up to 5 KK-levels in Eqs.\;\ref{dprime_sum} and \ref{eprime_sum} while estimating Wilson coefficients. The horizontal grey band corresponds to the 2$\sigma$ allowed range of experimental value of the above branching ratio.}
\label{nmUED}
\end{center}
\end{figure}

In this sub-section we present and discuss the main results of our analysis. We have already mentioned that in nmUED scenario KK-masses and various couplings among KK-excitations are the functions of BLT parameters. In Fig.\;\ref{nmUED} we have presented numerical values of branching ratio for $B\rightarrow X_s\gamma$ as function of BLT parameters. There are four different panels corresponding to four different values of  scaled gauge BLT parameter $R_V(\equiv r_V/R)$. In each panel, we show the variation of the branching ratio with $R^{-1}$ for different values  of scaled fermion BLT parameters $R_f(\equiv r_f/R)$.

Before going into  further details of numerical results, let us comment on the range of values of BLT parameters used in our analysis. Generically BLT parameters may positive or negative. However, it is clear from Eq.\;\ref{norm} that, for ${r_f}/{R}=-\pi$ the 0-mode solution becomes divergent and beyond  ${r_f}/{R} = - \pi$ the 0-mode fields become ghost-like.  
Hence any values of BLT parameters lower than $- \pi$ should be avoided. For the sake of completeness we have presented numerical results 
for some negative BLT parameters. Although, analysis of electroweak precision data \cite{bmm} disfavours large portion of negative BLT parameters.

Curves in each of the panels of Fig.\;\ref{nmUED} represent variation of branching ratios (SM+nmUED) of $B\rightarrow X_s\gamma$ with $R^{-1}$, for a specific set of BLT parameters ($R_V, R_f$). We have used the central value of Eq.\;\ref{nnlo} while calculating the branching ratio. The branching ratio increases with the increasing values of $R^{-1}$ and asymptotically converges to its SM value as $R^{-1}\rightarrow \infty$.  The suppression of the branching ratio for relatively smaller values of $R^{-1}$ is  a consequence of destructive interference of nmUED with the SM.

Furthermore, keeping $R_V$ fixed at positive values, if we change $R_f$ from positive to negative domain, KK-fermion masses would increase thus diminishing the values of loop functions $D'_n$ and $E'_n$. As in the concerned decay nmUED contribution destructively interferes with the SM contribution, decay branching ratio does not deviate drastically from its SM value. On the other hand, negative values of $R_V$ would also increase the KK-gauge boson masses, but in this case the couplings (via the overlap integrals $I^n_1$ and $I^n_2$) would make the nmUED contribution large leading to a large deviation of $B\rightarrow X_s\gamma$ branching ratio from its SM value.
If we contrast this limit on $R^{-1}$, with the same derived from the analysis of electroweak precision test (EWPT) \cite{bmm}, we see that EWPT would prefer higher values of $R^{-1}$ for negative  $R_f$. This can be 
explained from the enhanced coupling of a $n^{th}$ KK-W-boson to a pair of SM fermions \cite{bmm}\footnote{We must keep in mind that in nmUED, nontrivial correction to $T$ and $U$ parameters
arise via the correction of $G_F$ through $n^{th}$ KK-W-boson mediated 4-fermion interaction. For details see ref.\;\cite{bmm}.
}.

Comparing the branching ratio obtained from our calculation with experimental data, we could constrain the parameters of nmUED. We have three free parameters at our dispense, while we have only one experimental data (the branching ratio). So we would like to see what would be the lower limits on $R^{-1}$ for some fixed values of BLT parameters $R_f$ and $R_V$.  While deriving the limits we actually compare the quantity $-8.22\Delta C_7-1.99\Delta C_8 \equiv \delta $ with $2\sigma$ downward fluctuation of the difference between the experimental data (Eq.\;\ref{br_exp}) and the SM prediction  (Eq.\;\ref{br_sm}). We would like to remind that the above quantity ($\delta$) is exactly the 
nmUED contribution (including NNLO QCD corrections) to $B \rightarrow X_s \gamma$ branching ratio following  the Eq.\;\ref{nnlo}.  To obtain the 95\% C.L. lower limit on $R^{-1}$,
(for a fix set of values for $R_f$ and $R_V$) we add the experimental and theory errors in quadrature (call this quantity $\sigma$) and see for which value of $R^{-1}$, $\delta$ equals to the $2\sigma$  downward fluctuation  of the difference between SM and experimental values of the branching ratio. 

As for example, if $R_f = 2, \; R_V = 4$, $R^{-1} > 487 (503) \rm\;GeV$ when we sum unto 5(20) KK-levels. For $R_f = R_V = 4$, lower limit on 
$R^{-1}$ changes to $583 (600)  \rm\;GeV$. From the figures (Fig.\;\ref{nmUED}) it is clear that the limits are in the same ball park of those obtained from the analysis of $B_s \rightarrow \mu^+ \mu^-$ \cite{bmm}. For a more comprehensive list of  lower limits on $R^{-1}$ we refer the reader to Table \ref{sum-kk}. The numbers in the following table also shows that our results are not very sensitive to the number of KK-level considered in the sum while evaluating 
$\Delta C_7$ and $\Delta C_8$. The infinite series quickly converges as stated in \cite{gg_converge}.

\begin{table}[H]
\begin{center}
\hspace*{-1cm}
\resizebox{19cm}{!}{
\begin{tabular}{|c||c|c||c|c||c|c||c|c||c|c|}
\hline 
{}&\multicolumn{2}{|c||}{$R_V=-2$}&\multicolumn{2}{|c||}{$R_V=0$} &\multicolumn{2}{|c||}{$R_V=2$}&\multicolumn{2}{|c||}{$R_V=4$}&\multicolumn{2}{|c||}{$R_V=6$}\\
\hline{$R_f$}&
5 KK-level &  20 KK-level &
5 KK-level &  20 KK-level &
5 KK-level &  20 KK-level &
5 KK-level &  20 KK-level &
5 KK-level &  20 KK-level \\
\hline
-2&216.96&229.16&176.02&199.37&101.06&108.08&98.70&103.40&97.01&101.04\\
 0&148.59&159.58&323.16&342.15&273.10&288.31&269.77&285.97&266.47&283.63\\
 2&116.17&124.47&469.06&475.56&479.68&496.63&487.71&503.65&503.99&527.06\\
 4&102.95&108.08&544.95&548.12&573.83&587.22&583.22&599.62&596.66&613.66\\
 6&83.14&91.70&590.62&597.28&642.20&648.74&650.23&660.47&670.14&681.54\\
\hline
\end{tabular}
}
\end{center}
\caption[]{Lower limits on $R^{-1}$ (in GeV) derived from branching ratio of $B\rightarrow X_s\gamma$ for several values of BLT parameters showing the insensitivity on the number of  KK-levels in summation.}
\label{sum-kk}
\end{table}

\begin{figure}[t]
\begin{center}
\includegraphics[scale=1.3,angle=0]{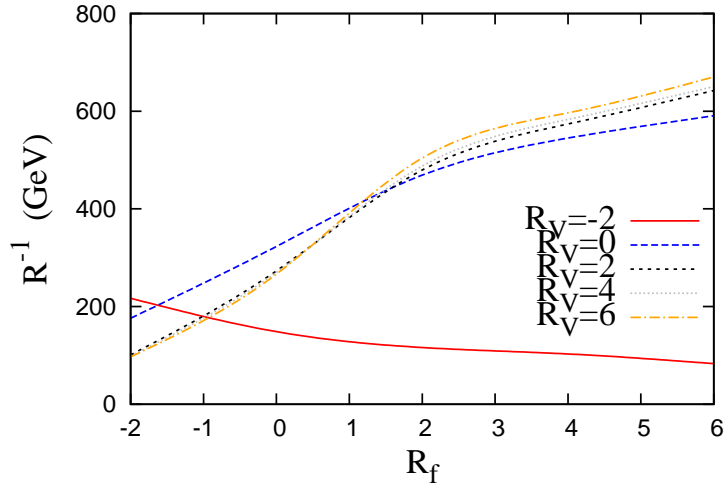}
\caption{95\% C.L. exclusion contours in $R_f - R^{-1}$ plane for five different choices of $R_V$ from branching ratio of $B\rightarrow X_s\gamma$ decay. The area below a particular line (fixed $R_V$) has been excluded at 95\% C.L.}
\label{nmUED_band}
\end{center}
\end{figure}

Finally, in Fig.\;\ref{nmUED_band} we would like to show the region of parameter space which has been excluded by current experimentally measured branching ratio of $B\rightarrow X_s\gamma$.  In this figure we have plotted contours  corresponding to five different  values of $R_V$ in $R_f-R^{-1}$ plane. The region below a particular line  has been excluded at 95\% C.L.  Nature of these contours can be understood with the help of Fig.\;\ref{nmUED}. Let us focus on a particular panel which corresponds to a  positive value of $R_V$. As we have seen that negative BLT parameters would not give any meaningful limit on $R^{-1}$, so we have restricted our discussions for positive BLT parameters only. If we imagine a vertical line corresponding to a fixed value of $R^{-1}$, then we can see that the branching ratio is lower for higher values of $R_f$, which is obvious because KK-masses decrease with increasing BLT parameters. We have already mentioned that KK-contributions destructively interfere with SM contributions, hence for lower KK-mass we obtain lower branching ratio. So to obtain a fixed value of branching ratio at  higher values of $R^{-1}$, we need higher values of $R_f$. Overall this effect is slightly enhanced by higher values of $R_V$.

To this end, we would like to mention that we have restricted ourselves for the choice of BLT parameters up to 6. The reason is that, if we look at the Fig.\;\ref{nmUED_band} (or Table\;\ref{sum-kk}) then it is clear that the lower limits on $R^{-1}$ are weakly sensitive to the
gauge BLT parameter $R_V$. So, we expect that lower limit on $R^{-1}$  for greater values of $R_V$ would not drastically change.

$\bullet$ {\bf Bound on {\boldmath$R^{-1}$} in UED model}

Before we conclude let us quickly review the impact our analysis on UED model. Branching ratio of  $B\rightarrow X_s\gamma$ in the UED model can be straightforwardly obtained from the nmUED results if we set $r_V = r_f =0$. Under this situation the functions $D'_n(x_t,x_{f^{(n)}},x_{V^{(n)}})$ and $E'_n(x_t,x_{f^{(n)}},x_{V^{(n)}})$ given in Eqs.\;\ref{dprime} and \ref{eprime} would convert into their UED 
forms. Further, in this limit $n^{th}$ mode of KK-mass become equal to  $nR^{-1}$ while the overlap integrals $I^n_1$ and $I^n_2$ become unity. We have verified that in this vanishing BLT limit the forms of these functions ($D'_n(x_t,x_{f^{(n)}},x_{V^{(n)}})$ and $E'_n(x_t,x_{f^{(n)}},x_{V^{(n)}})$) are identical with the one in ref.\;\cite{buras3}\footnote{In ref.\;\cite{buras3} mass of an $n^{th}$ KK-excitation is $nR^{-1}$, as the authors of this article \cite{buras3} have not considered any radiative corrections to the KK-masses in their analysis.}. One can easily read the lower limit on $R^{-1}$ to be 323 GeV for $r_V = r_f =0$ from Table 1.
 Now this results is quite compatible with those that obtained from several other processes. For example $(g-2)_\mu$ \cite{mu2}, $\rho$-parameter \cite{rho}, FCNC process \cite{buras3, buras, fcnc} and electroweak observables like $R_b$ \cite {zbb, elctrowk} would result into a lower limit on $R^{-1}$ which is in the ballpark of 300 GeV. However,  the projected tri-lepton signal at 8 TeV LHC one can derive lower limit on $R^{-1}$ up to 1.2 TeV\cite{belayev}, \cite{Working_group}.

Let us briefly compare our results with that obtained in ref.\;\cite{bsg}. In ref.\;\cite{bsg} the authors have calculated Wilson coefficients at the electroweak scale at the LO in UED model. They have not considered any further higher order corrections to this process. Subsequently they have compared their theoretical result with  SM prediction which includes NLO QCD corrections. They have obtained a lower bound on $R^{-1}(> 600$ GeV) at 95\% C.L..  Incorporation of NNLO correction to our result reduce the effect of destructive interference  in the branching ratio. That is to say, NNLO correction pushes the UED results more towards the experimental number, thereby reducing the value of lower limit on $R^{-1}$ in our case in comparison to \cite{bsg}.
 

\section{Conclusion}
We have computed the effects of KK-excitations to the branching ratio of $B\rightarrow X_s\gamma$ in a $4 + 1$ dimensional scenario, called non-minimal Universal Extra Dimensional model, where all SM fields can propagate in the extra spatial dimension. This model is hallmarked by several boundary localised terms (kinetic, Yukawa etc.). Coefficients of these boundary localised terms parametrise the unknown radiative corrections to the masses and couplings in the full 5D theory. Consequently couplings and mass spectrum of KK-modes are modified in a non-trivial manner in the 4D effective theory in comparison to the minimal UED model. In our analysis we have used two different classes of BLT parameters. One is $r_V$ which  specifies coefficients of boundary terms for the gauge and Higgs sectors while $r_f$ represents coefficients of boundary terms of fermions and Yukawa interactions. We have studied the effects of these BLT parameters on $B\rightarrow X_s\gamma$ decay process.   

Effective Hamiltonian for the decay process $B\rightarrow X_s\gamma$ can be parametrised by series of Wilson coefficients. Among these coefficient we have calculated the coefficients for electromagnetic and chromomagnetic dipole operators. The corresponding Feynman (penguin) diagrams are listed in Fig.\;\ref{magnetic_pen}. Exploiting GIM mechanism we have incorporated contributions from 3 generations of quarks in our analysis. We estimate the total contribution coming from the penguin diagrams we have also considered the SM ($0^{th}$ KK-mode) contributions with the KK-contributions. In view of a recent analysis relating the Higgs boson mass and cut-off of a UED theory \cite{kksum} we summed up to 5 KK-levels in our calculation. And finally using the ref.\;\cite{NNLO} we have incorporated NNLO QCD corrections in our analysis. 

Present experimental data and theory prediction for $Br(B_s \rightarrow X_s \gamma$) allow  only a narrow window for any BSM contribution to this decay amplitude. We have constrained the parameter space of nmUED model using the experimental data . Moreover in the vanishing BLT limits, we can reproduce the results of UED model ($R_f = R_V = 0$ situation) from our calculation. Using the  
present analysis we obtained a lower limit of 323 GeV on $R^{-1}$  at 95 \% C.L. This limit on $R^{-1}$ in UED model, is in the same ballpark with the limits those are obtained from the consideration of $R_b$ \cite{zbb}, $\rho$-parameters \cite{rho} or $B_s \rightarrow  \mu^+ \mu^-$ \cite{bmm}. As in the nmUED model apart from the compactification radius there exists two extra BLT parameters  (strictly holds for our consideration) so the bounds of the lower limit on $R^{-1}$ 
in nmUED model would be much more relaxed than the UED model. Hence depending on the values BLT parameters (for example $R_V=6$ and $R_f=6$) the range of lower limit on $R^{-1}$ can be as high as  670 GeV. This value is definitely very promising in nmUED model. Thus the recent experimental result of the branching ratio of $B\rightarrow X_s \gamma$ can exclude large portion of the parameters space of this nmUED scenario. It is obvious from Fig.\;\ref{nmUED_band} that lower limits on $R^{-1}$ for negative 
values of BLT parameters are not so competitive and these values have been already ruled out from the consideration of electroweak precision data \cite{bmm}.

{\bf Acknowledgements} AD is partially supported by  Department of Science and Technology, Science and Engineering Research Board (DST-SERB) research project.  AS acknowledges  support from Regional Centre for Accelerator-based Particle
Physics (RECAPP), Harish-Chandra Research Institute, where the major portion of this work has been performed. Authors are grateful to Gautam Bhattacharyya for useful discussion.

\begin{appendices}
\renewcommand{\thesection}{\Alph{section}}
\renewcommand{\theequation}{\thesection-\arabic{equation}} 

\setcounter{equation}{0}

\section{Feynman rules for \boldmath{$B\rightarrow X_s\gamma$} in nmUED}\label{fyerul}
In this appendix we have listed the necessary Feynman rules needed for our computations. Assuming all momenta and fields are assumed to be incoming.  

1) $A^{\mu}W^{\nu\pm}S^{\mp}$
$\displaystyle : {g_2}{s_w M_{W^{(n)}}} g_{\mu\nu} C$, where $C$ is given by:

\begin{equation}
 \begin{aligned}
  A^{\mu} W^{\nu(n)+} G^{(n)-}: C   &= 1,\\ 
  A^{\mu} W^{\nu(n)-} G^{(n)+}: C   &= -1,\\
  A^{\mu} W^{\nu(n)+} H^{(n)-}: C   &= 0,\\
  A^{\mu} W^{\nu(n)-} H^{(n)+}: C   &= 0.  
 \end{aligned}
\end{equation}

Here $g_2$ is the $SU(2)$ gauge coupling constant and $s_w$ is the $\sin$ of Weinberg angle ($\theta_w$).
\newpage
2) $A^{\mu}S^{\pm}_1S^{\mp}_2$
$\displaystyle : -{ig_2}{s_w} (k_2-k_1)_{\mu} C$, where $C$ is given by:

\begin{equation}
 \begin{aligned}
  A^{\mu} G^{(n)+} G^{(n)-}: C &= 1,\\   
  A^{\mu} H^{(n)+} H^{(n)-}: C &= 1,\\
  A^{\mu} G^{(n)+} H^{(n)-}: C &= 0,\\ 
  A^{\mu} G^{(n)-} H^{(n)+}: C &= 0.
\end{aligned}
\end{equation}

Here the scalar fields $S\equiv H,G.$

3) $A^{\mu}(k_1)W^{\nu+}(k_2)W^{\lambda-}(k_3)$
$\displaystyle :$
\begin{equation}
 ig_2s_w \left[ g_{\mu\nu} (k_2
      -k_1)_\lambda + g_{\mu\lambda} (k_1 -k_3)_\nu +
    g_{\lambda\nu} (k_3 -k_2)_\mu \right].
\end{equation}

4) $A^{\mu}{\overline{f}_1} f_2$
  $\displaystyle  : {i g_2}{s_w} \gamma_\mu C$, where $C$ is given by:

\begin{equation}
 \begin{aligned}
  A^{\mu} \bar{u_i} u_i: C &= \frac23,\\   
  A^{\mu} {\overline{T}^{1(n)}_i} T^{1(n)}_i: C &= \frac23,\\
  A^{\mu} {\overline{T}^{2(n)}_i} T^{2(n)}_i: C &= \frac23,\\ 
  A^{\mu} {\overline{T}^{1(n)}_i} T^{2(n)}_i: C &= 0,\\
  A^{\mu} {\overline{T}^{2(n)}_i} T^{2(n)}_i: C &= 0.
\end{aligned}
\end{equation}

5) $G^{\mu}{\overline{f}_1} f_2$
  $\displaystyle  : {i g_s}{T^a_{\alpha\beta}} \gamma_\mu C$, where $C$ is given by:

\begin{equation}
 \begin{aligned}
  G^{\mu} \bar{u_i} u_i: C &= 1,\\   
  G^{\mu} {\overline{T}^{1(n)}_i} T^{1(n)}_i: C &= 1,\\
  G^{\mu} {\overline{T}^{2(n)}_i} T^{2(n)}_i: C &= 1,\\ 
  G^{\mu} {\overline{T}^{1(n)}_i} T^{2(n)}_i: C &= 0,\\
  G^{\mu} {\overline{T}^{2(n)}_i} T^{2(n)}_i: C &= 0,
\end{aligned}
\end{equation}

6) $S^{\pm}{\overline{f}_1} f_2$
  $\displaystyle  = \frac{g_2}{\sqrt{2} M_{W^{(n)}}} (P_L C_L + P_R C_R)$, where $C_L$ and $C_R$ are given by:

\begin{equation}
\begin{aligned}
  & G^+ \bar{u_i} d_j :  &
  &\left\{\begin{array}{l}C_L = -m_i V_{ij},\\
      C_R = m_j V_{ij},\end{array}\right.
  &&G^- \bar{d_j} u_i :     &
  &\left\{\begin{array}{l}C_L = -m_j V_{ij}^*,\\
      C_R = m_i V_{ij}^*,\end{array}\right.\\
  & G^{(n)+}{\overline{T}^{1(n)}_i} d_j :  &
  &\left\{\begin{array}{l}C_L = -m_1^{(i)} V_{ij},\\
      C_R = M_1^{(i,j)} V_{ij},\end{array}\right.
  &&G^{(n)-}\bar{d_j}T^{1(n)}_i :   &
  &\left\{\begin{array}{l}C_L = -M_1^{(i,j)} V_{ij}^*,\\
     C_R = m_1^{(i)} V_{ij}^*,\end{array}\right.\\
  & G^{(n)+}{\overline{T}^{2(n)}_i} d_j :  &
  &\left\{\begin{array}{l}C_L = m_2^{(i)} V_{ij},\\
      C_R =-M_2^{(i,j)} V_{ij},\end{array}\right.
  &&G^{(n)-}\bar{d_j}T^{2(n)}_i :   &
  &\left\{\begin{array}{l}C_L = M_2^{(i,j)} V_{ij}^*,\\
     C_R =-m_2^{(i)} V_{ij}^*,\end{array}\right.\\
  & H^{(n)+}{\overline{T}^{1(n)}_i} d_j :  &
  &\left\{\begin{array}{l}C_L = -m_3^{(i)} V_{ij},\\
      C_R = M_3^{(i,j)} V_{ij},\end{array}\right.
  &&H^{(n)-}\bar{d_j}T^{1(n)}_i :   &
  &\left\{\begin{array}{l}C_L = -M_3^{(i,j)} V_{ij}^*,\\
     C_R = m_3^{(i)} V_{ij}^*,\end{array}\right.\\
  & H^{(n)+}{\overline{T}^{2(n)}_i} d_j :  &
  &\left\{\begin{array}{l}C_L = m_4^{(i)} V_{ij},\\
      C_R =-M_4^{(i,j)} V_{ij},\end{array}\right.
  &&H^{(n)-}\bar{d_j}T^{2(n)}_i :   &
  &\left\{\begin{array}{l}C_L = M_4^{(i,j)} V_{ij}^*,\\
     C_R =-m_4^{(i)} V_{ij}^*,\end{array}\right.
\end{aligned}
\end{equation}


7) $W^{\mu\pm}{\overline{f}_1}f_2$
  $\displaystyle  :  \frac{i g_2}{\sqrt{2}} \gamma_\mu P_L C_L$, where $C_L$ is given by:

\begin{equation}
\begin{aligned}
  & W^{\mu+}\bar{u_i} d_j : &&     C_L = V_{ij},
  && W^{\mu-}\bar{d_j} u_i : &&    C_L = V^*_{ij},\\
  & W^{\mu(n)+}{\overline{T}^{1(n)}_i}d_j : &&   C_L = I^n_1\;c_{in} V_{ij},
  &&W^{\mu(n)-}\bar{d_j}{{T}^{1(n)}_i} : && C_L = I^n_1\;c_{in} V^*_{ij},\\
  & W^{\mu(n)+}{\overline{T}^{2(n)}_i}d_j : &&   C_L = -I^n_1\;s_{in} V_{ij},
  &&W^{\mu(n)-}\bar{d_j}{{T}^{2(n)}_i} : && C_L = -I^n_1\;s_{in}V^*_{ij}.
\end{aligned}
\end{equation}

Here the fermion fields $f\equiv u, d, T^1_t, T^2_t.$

The mass parameters $m_x^{(i)}$ are given by \cite{bmm}:
\begin{equation}
\label{mparameters}
  \begin{aligned}
    m_1^{(i)} &= I^n_2\;m_{V^{(n)}}c_{in} +I^n_1\;m_i s_{in},\\
    m_2^{(i)} &= -I^n_2\;m_{V^{(n)}}s_{in}+I^n_1\;m_i c_{in},\\
    m_3^{(i)} &= -I^n_2\;iM_W c_{in} +I^n_1\;i\frac{m_{V^{(n)}}m_i}{M_W}s_{in},\\
    m_4^{(i)} &= I^n_2\;iM_W s_{in}+I^n_1\;i\frac{m_{V^{(n)}}m_i}{M_W}c_{in},
  \end{aligned}
\end{equation}
where $m_i$ represents the mass of the 0-mode {\it up-type} fermion and $c_{in}=\cos(\alpha_{in})$ and $s_{in}=\sin(\alpha_{in})$ with $\alpha_{in}$ as defined earlier.

And the mass parameters $M_x^{(i,j)}$ are \cite{bmm}:
\begin{equation}\label{Mparameters}
  \begin{aligned}
    M_1^{(i,j)}  &= I^n_1\;m_j c_{in},\\
    M_2^{(i,j)}  &= I^n_1\;m_j s_{in},\\
    M_3^{(i,j)}  &= I^n_1\;i\frac{m_{V^{(n)}}m_j}{M_W}c_{in},\\
    M_4^{(i,j)}  &= I^n_1\;i\frac{m_{V^{(n)}}m_j}{M_W}s_{in},
  \end{aligned}
\end{equation}
where $m_j$ represents the mass of the 0-mode {\it down-type} fermion. Here, $I^n_1$ and $I^n_2$ are the overlap integrals are given in the following \cite{bmm}:

\begin{equation}
I^n_1 = 2\sqrt{\frac{1+\frac{r_V}{\pi R}}{1+\frac{r_f}{\pi R}}}\left[ \frac{1}{\sqrt{1 + \frac{r^2_f m^2_{f^{(n)}}}{4} + \frac{r_f}{\pi R}}}\right]\left[ \frac{1}{\sqrt{1 + \frac{r^2_V m^2_{V^{(n)}}}{4} + \frac{r_V}{\pi R}}}\right]\frac{m^2_{V^{(n)}}}{\left(m^2_{V^{(n)}} - m^2_{f^{(n)}}\right)}\frac{\left(r_{f} - r_{V}\right)}{\pi R},
\label{i1}
\end{equation}

\begin{equation}
I^n_2 = 2\sqrt{\frac{1+\frac{r_V}{\pi R}}{1+\frac{r_f}{\pi R}}}\left[ \frac{1}{\sqrt{1 + \frac{r^2_f m^2_{f^{(n)}}}{4} + \frac{r_f}{\pi R}}}\right]\left[ \frac{1}{\sqrt{1 + \frac{r^2_V m^2_{V^{(n)}}}{4} + \frac{r_V}{\pi R}}}\right]\frac{m_{V^{(n)}}m_{f^{(n)}}}{\left(m^2_{V^{(n)}} - m^2_{f^{(n)}}\right)}\frac{\left(r_{f} - r_{V}\right)}{\pi R}.
\label{i2}
\end{equation}

The characteristics dependence of these integrals on BLT parameters has been illustrated in \cite{bmm} with conjunction of two figures (see Fig.~1 of ref.\;\cite{bmm} and the corresponding discussions). One can easily check that how these integrals affect the interactions which are involved in our analysis.

\end{appendices}

\end{document}